\input harvmac
\def \la{\longrightarrow}

\def \ha  { {\textstyle{1\ov 2} } }

\def \a {\alpha}


\def\ys{\bar s}
\def\yt{\bar t}
\def\yu{\bar u}
\def\us{s_{pq}}
\def\ut{t_{pq}}
\def\uu{u_{pq}}
\def\asl{A_4^{sl(2)}(s,t,u)}

\def\sld{$ SL(2,{\bf Z})$}


\def \td {\tilde }

\def \ov {\over }
\def \four{{\textstyle{1\over 4}}}

\def \lr { \lref}
\def\np {{  Nucl. Phys. }}
\def \pl {{  Phys. Lett. }}

\def \ijmp {{ Int. J. Mod. Phys. }}

\baselineskip8pt
\Title{
\vbox
{\baselineskip 6pt{\hbox{ }}{\hbox
{Imperial/TP/96-97/57}}{\hbox{hep-th/9707241}} {\hbox{
  }}} }
{\vbox{\centerline {An ansatz  for a non-perturbative four-graviton}
\vskip4pt
 \centerline {amplitude in type IIB superstring theory}
}}
\vskip -27 true pt
\centerline  {  Jorge G. Russo{\footnote {$^*$} {e-mail address:
j.russo@ic.ac.uk
 } } 
}

\bigskip

 \centerline {\it  Theoretical Physics Group, Blackett Laboratory,}
\smallskip
\centerline {\it  Imperial College,  London SW7 2BZ, U.K. }

\medskip\bigskip

\centerline {\bf Abstract}
\medskip
\baselineskip10pt
\noindent
A natural \sld -invariant generalization of the Veneziano amplitude in type IIB
superstring theory is investigated. It includes certain perturbative and 
non-perturbative (D-instanton) contributions, and it reduces to the correct 
expressions
in different limits. The singularities are  poles in the $s$-$t$-$u$ 
channels,
corresponding to the exchange of particles with a mass spectrum coinciding
with that of  $(p,q)$ string states.
We describe the general structure of the associated
perturbative corrections to the effective action.

\medskip
\Date {July 1997}
\noblackbox
\baselineskip 14pt plus 2pt minus 2pt

\lr\hult{
C.M. Hull and P.K. Townsend, Nucl. Phys. { B438} (1995) 109.}

\lr \green{M.B. Green and M. Gutperle, hep-th/9604091.}

\lr \townelev{ P.K. Townsend, \pl B350 (1995) 184, hep-th/9501068.  }

\lr \witten {E. Witten, \np B460 (1995) 335.}

\lr \tow {P.K. Townsend,  hep-th/9609217.} 

\lr \schwa  {For reviews and further references see: J.H. Schwarz,
hep-th/9607201;
 M.J. Duff, hep-th/9608117;  P.K. Townsend,  hep-th/9609217.}
 
\lr \bergsh{ E. Bergshoeff, E.  Sezgin and P.K. Townsend, \pl B189 (1987)
75.}

\lr\banks{
T. Banks, W. Fischler, S.H. Shenker and L. Susskind, 
   hep-th/9610043.}
\lr \tser{A.A. Tseytlin, hep-th/9609212. }
\lr\russo {J.G. Russo, 
hep-th/9610018 .}
\lr \john{J.H. Schwarz, \pl B360 (1995) 13 
(Erratum-ibid.B364 (1995) 252), hep-th/9508143.}

\lr\rusty{J.G. Russo and A.A. Tseytlin, \np B490 (1997) 121;
J.G. Russo, hep-th/9703118.}
\lr \gsh {M.B. Green  and J.H. Schwarz, \np B198 (1982) 441.}
\lr \gsb {
 M.B. Green, J.H. Schwarz and L. Brink,  \np B198 (1982) 474. }
\lr \mets { R.R. Metsaev and A.A.  Tseytlin, \np B298 (1988) 109. }
\lr \frats {E.S. Fradkin   and A.A. Tseytlin, \np B227 (1983) 252.}
\lr \duftom{ M.J. Duff and  D.J. Toms, 
in:  {\it Unification of the Fundamental 
Particle Interactions II},  Proceedings of the Europhysics Study Conference, 
Erice, 4-16 October 1981, 
ed. by J. Ellis and S. Ferrara (Plenum, 1983).} 
\lr \kallosh { R. Kallosh,  unpublished; P. Howe, unpublished. }
\lr \gris {M. Grisaru and W. Siegel,  \np B201 (1982) 292.}
\lr \sund {B. Sundborg,  \np B306 (1988) 545.}
\lr \acv { D. Amati, M. Ciafaloni and G. Veneziano, \ijmp  3 (1988) 615;
 \np B347 (1990) 550.}
\lr \lip { L.N. Lipatov, \pl B116 (1982) 411; \np B365 (1991) 614.}
\lr \cia {  M. Ademollo, A. Bellini and M. Ciafaloni, \np B393 (1993) 79.}
\lr \grisa{M.T. Grisaru, A.E.M. van de Ven and D. Zanon, \np B277 (1986) 388, 409.}
\lr \grow{D.J. Gross and E. Witten, \np B277 (1986) 1.}

\lr \bbpt{ K. Becker, M. Becker, J. Polchinski and A.A. Tseytlin, 
hep-th/9706072.   } 
\lr \sak { N. Sakai and Y. Tanii, \np B287 (1987) 457.}

\lr \witt { E. Witten, \np B443 (1995) 85, hep-th/9503124.}
\lr \ggv { M.B. Green, M.  Gutperle and P.  Vanhove, hep-th/9706175. }

\lr \polch {J. Polchinski, {\it  TASI Lectures on D-branes},
hep-th/9611050.}
 
\lr \gv {M.B. Green and P. Vanhove, hep-th/9704145.}

\lr \grgu{M.B. Green and M. Gutperle, hep-th/9701093.}
\lr \tte{A.A. Tseytlin, \np B467 (1996) 383, hep-th/9512081.}
\lr\kiri{E. Kiritsis and B. Pioline, hep-th/9707018.}
\lr\rutse{J.G. Russo and A.A. Tseytlin, hep-th/9707134.}
\lr\anto{I. Antoniadis, B. Pioline and T.R. Taylor, hep-th/9707222.}
\lr\towns{P.K. Townsend, hep-th/9705160; M.~Cederwall and
P.K.~Townsend, hep-th/9709002.}

\lr \tte{A.A. Tseytlin, \np B467 (1996) 383, hep-th/9512081.}
\lr \ber{ M. Bershadsky, S. Cecotti, H. Ooguri and C. Vafa, Commun. Math. Phys.
165 (1994) 31,  hep-th/9309140.}
\lr \anton{
 I. Antoniadis, E. Gava, K.S. Narain and T.R. Taylor, \np B455 (1995) 109,
hep-th/9507115.}
\lr\russo{For a recent review see: J.G. Russo, hep-th/9703118.}

\def\aa{ {\cal V } }


\def \L { \Lambda_{11}}
\def \RR {{\cal R}}

\def\slz{ $SL(2,{\bf Z})$}
\def\brs{\bar s}
\def\brt{\bar t}
\def\bru{\bar u}


A number of interesting  aspects of M-theory have recently been clarified
\refs {\witt, \schwa }. Its fundamental
degrees of freedom are not yet completely understood, and
an adequate  formulation is  still elusive.
Such  degrees of freedom could be revealed
if an exact scattering amplitude is determined.
In conventional quantum theories, scattering amplitudes
exhibit, through their analytic structure,
the physical states of the theory (appearing as poles in the different channels),
and the residue of the poles gives
important information about the couplings. 
In M-theory, the simplest non-trivial process that one can investigate 
is the four-graviton amplitude.
Some progress in this direction was made in refs. \refs{\grgu  \gv \ggv
-\rutse  }. In ref. \grgu , an exact function of the coupling
multiplying the $\RR ^4$ term in type II superstring theories was proposed.
Such function,
which was further motivated in refs. \gv\ and \ggv ,
 represents the leading term in a low-energy 
expansion of the four-graviton scattering amplitude.
In ref. \ggv\ it was shown that this function can also be obtained from
the one-loop four graviton amplitude in eleven dimensional supergravity
with space-time topology ${\bf R}^{9}\times T^2$.
The  one-loop graviton amplitude including all powers of external momenta
 was then calculated in \rutse . 


In this paper we shall concentrate on ten-dimensional
type IIB superstring theory. A generalization of the present result to type 
IIB on
${\bf R}^{9}\times S^1 $ would allow, by string dualities,
 the determination of the corresponding four-graviton amplitude in M-theory.
 Here the basic idea is to exploit the \slz\ symmetry of 
 type IIB superstring theory, which requires that
the effective action must be invariant under $SL(2,{\bf Z})$ transformations
to all orders in the $\a' $ expansion. 
In the Einstein frame, a term of given order in derivatives
involving the metric 
must be multiplied by a modular  function of the coupling,
as happens in the $\RR^4$ example investigated in \refs{\grgu, \gv}.
Since there is a one-to-one correspondence between 
certain (\slz\ invariant) terms in the effective action (see eq. (20)~) and the terms of the
momentum expansion
of the four-graviton amplitude, the same modular  functions
 appear in the four-graviton amplitude, which
must therefore be invariant under $SL(2,{\bf Z})$ transformations. 
This symmetry property constrains its allowed structure.
 It will be seen that there is a simple \slz -invariant scattering amplitude 
 that in both  weak coupling ($g_B\ll 1$) and low-energy 
($\a' {\rm (momentum)}^2 \ll 1$)
regimes reproduces the known expressions. Its analytic structure 
is  consistent with
the expected particle spectrum.  
 
Our starting point is the tree-level scattering amplitude:
\eqn\venez{
A_4=\kappa^2  K A_4^0(s,t,u)\ ,
}
\eqn\ven{
A_4^0(s,t,u)={1\ov \bar s\bar t \bar u} {\Gamma (1- \bar s)\Gamma (1-\bar t)
\Gamma(1-\bar u)\ov \Gamma (1+ \bar s)\Gamma (1+\bar t)
\Gamma(1+\bar u) }\ ,
}
$$
\bar s=\four \a' s\ ,\  \ \ \bar t=\four \a' t\ , \ \ \ \bar u=\four
\a' u\ ,\ \ \ \ \ys+\yt+\yu =0\ ,
$$
where $K$ is the usual kinematic factor \grow , depending on the momenta and 
polarization of the external
states ($K\sim ({\rm momentum})^8$).
We will find convenient to work with the logarithm of the amplitude,
which has a simple expansion in powers of $\bar s,\bar t, \bar u$, i.e.
\eqn\logve{
\ln A_4^0(s,t,u)=-\ln \brs\brt\bru +2\sum_{k=1}^{\infty }
{\zeta (2k+1)\ov 2k+1} \big(\brs ^{2k+1}+\brt ^{2k+1}+\bru ^{2k+1} \big)\ ,
}
or
\eqn\aaa{
A_4^0(s,t,u)={1\ov \brs\brt\bru }+  2 \zeta(3)+2\zeta(5) 
\big( \brs ^2+\brs\brt+\brt^2 \big)+...\ 
}
The term $\zeta (3)\RR^4 $ in the effective action, associated with the
second ($k=1$) term in the above expansion, 
has been the subject of
the investigations in \refs{\grgu -\ggv }.
In these works, it was argued that the exact function of the coupling
 multiplying $\RR^4 $ (which incorporates all perturbative and non-perturbative
 contributions) is essentially obtained by the substitution
$$
\zeta (3)=\sum _{m=1}^\infty {1\ov m^3} \ \la\ \ha \sum_{(m,n)\neq (0,0)}
{1\ov |m+n\tau |^3}=\zeta (3) \tau _2 ^{-3/2} E_{3/2}(\tau )\ ,
$$
where $\tau =\tau_1+i\tau_2 =C^{(0)}+i g^{-1}_B $ 
is the usual coupling of type IIB superstring theory and
$E_r(\tau )$ is the generalized Einsestein series (sometimes called
Epstein zeta-function)
\eqn\eins{
E_r(\tau )=
\sum_{(p,q)'} {\tau_2^r\ov |p+q \tau |^{2 r}}\ .
}
The notation $(p,q)'$ means that $p$ and $q$ are relatively prime.
A natural question is then whether a similar substitution for the
next $\zeta(2k+1)$ terms in eq. \logve\ would also account 
for   perturbative and non-perturbative
contributions. A further motivation for this guess is that
modular functions $E_{k+1/2} (\tau )$ have also appeared in the one-loop
four-graviton scattering amplitude of eleven dimensional supergravity
on ${\bf R}^9\times T^2$ calculated in \rutse , and they indeed account for
contributions of Kaluza-Klein states. 
Thus we suggest the following  generalization
of the Veneziano amplitude: 
$$
A_4=\kappa^2  K \asl \ ,
$$
with
\eqn\vex{
\ln A_4^{ sl(2)}(s,t,u)=-\ln \brs\brt\bru +2 \sum_{k=1}^{\infty }
{\zeta (2k+1)g_B^{k+1/2} E_{k+1/2}(\tau )\ov 2k+1} \big(\brs ^{2k+1}
+\brt ^{2k+1}+\bru ^{2k+1} \big)\ .
}
The \slz\ invariance is  explicit in the Einstein frame, 
$g_{\mu\nu}^E=g_B^{-1/2} g_{\mu\nu}$, so that $s_E=g_B^{1/2} s$, etc.
As a result, the
factor  $g_B^{k+1/2}$ in the numerator of eq.~\vex\  is absorbed into the
Einstein-frame variables $s_E,t_E,u_E$. The only  dependence on the coupling that remains
is in the invariant modular functions $E_{k+1/2}(\tau )$.

Using eq.~\eins , the amplitude \vex\ can be written as (cf. eq.~\logve\ )
\eqn\vexx{
\ln A_4^{ sl(2)}(s,t,u)=-\ln \brs\brt\bru +2 \sum_{(p,q)'} \sum_{k=1}^{\infty }
{\zeta(2k+1)\ov 2k+1} \big(\us ^{2k+1}
+\ut ^{2k+1}+\uu ^{2k+1} \big)\ ,\ \ \ 
}
\eqn\stuu{
\us ={\a' s\ov 4|p+q\tau |}\ ,\ \ \ \ut ={\a' t\ov 4|p+q\tau |}\ ,\ \ \ 
\uu ={\a' u\ov 4|p+q\tau |}\ ,\ \ \ \ \us+\ut+\uu =0\ .
}
Thus
\eqn\vvx{
\asl ={1\ov \bar s\bar t \bar u} \prod_{(p,q)'}
{\Gamma (1- \us)\Gamma (1-\ut)
\Gamma(1-\uu )\ov \Gamma (1+ \us )\Gamma (1+\ut )
\Gamma(1+\uu ) }\ .
}
By expressing  the $\zeta $-function in eq.~\vexx\ as a sum, the amplitude can alternatively be written 
in the form
\eqn\smn{
\asl =-{1\ov \bar s\bar t\bar u}\prod_{ m\geq 0, n}
{ (|m+n\tau|+\bar s)(|m+n\tau|+\bar t)(|m+n\tau |+\bar u)\ov
  (|m+n\tau|-\bar s)(|m+n\tau|-\bar t)(|m+n\tau |-\bar u)}\ .
}
Let us summarize the properties of the amplitude $A_4=\kappa^2  K \asl $ :
\smallskip
\noindent i) It reduces to the Veneziano amplitude \venez\ at weak coupling,
 $g_B \to 0$ (see also eq.~(18)~).

\noindent ii) In the 
low-energy limit, it reduces  to the massless pole plus 
a $\zeta(3)E_{3/2}(\tau )$ term (see eq.~\vex\ ).

\noindent iii) It is invariant under \slz\ transformations.

\noindent iv) It has correct perturbative ($g^{2k}_B$) and
nonperturbative $O(e^{-2\pi k/g_B})$ dependence on the string coupling 
(see eqs.~(15),~(16)~).

\noindent v) It has poles in the $s$-$t$-$u$ channels at $\us=-n $, $\ut=-n $,
$\uu=-n $, $n=0,1,2,...$
corresponding to exchange of particles with masses
\eqn\polos{
\four \a' M^2 =n |p+q\tau |\ .
}
This spectrum exactly  coincides with  the  spectrum of  $(p,q)$ string states 
\john:
\eqn\ppqq{
 M^2=4\pi T_{pq}(N_R+N_L)={2\ov\a'} \ |p+q\tau |\ (N_R+N_L)\ ,\ \ \ \ N_R=N_L\ .
}
The string tension $T_{pq}=T\ |p+q\tau|$ was originally derived
for winding strings of type IIB compactified on the circle.
Equation \ppqq\ corresponds to the zero-winding sector of the
spectrum of  ref.~\john\ (i.e., it describes states with vanishing NS-NS and R-R charges).
It is  actually not clear why the spectrum \ppqq\ 
should apply at any coupling,
since the masses of these states do not seem to be protected by supersymmetry 
in an obvious way (except for $N_L=N_R=0$).
Once all quantum corrections have been taken into account, 
poles corresponding to  unstable particles  should lie away from the real axes.
Moreover, there should appear  cuts as required by unitarity, associated
with intermediate states.\foot{
I thank M. Green for valuable comments on this point.}
The amplitude \vvx , which has no cuts, may represent the 
``tree-level" result  of an underlying $SL(2,{\bf Z})$-invariant 
perturbative expansion.\foot{
Progress in this direction was recently made in ref.~\towns ,
where a manifest S-dual action for the type IIB superstring was introduced.}

Such a remarkable re-organization of higher genus and non-perturbative 
corrections
may be automatic in M-theory compactified on a 2-torus. 
Indeed, from the eleven-dimensional viewpoint, the $SL(2,{\bf Z})$
symmetry is a simple 
consequence of the reparametrization symmetry of the 2-torus.
The ten-dimensional limit of the type IIB superstring corresponds
to a small torus area, $R_{10},R_{11} \to 0$,
keeping the ratio $R_{11}/R_{10}$ finite. 
In this limit the \slz\ symmetry
of the amplitude is preserved.
A concrete example is provided by eleven-dimensional supergravity
on ${\bf R}^9\times T^2$.
The one-loop supergravity amplitude on ${\bf R}^9\times T^2$ 
is given by~\rutse
$$
\big( 2\pi g_B^2R_{10}\big)^{-1} \ A_{4\rm T} (s,t)= 
   {2\ov 3 } \aa \L^3+ { \zeta (3) E_{{3/2}}(\tau )\ov \pi \aa ^{1/2} }
   - 2\sqrt{\pi }  s^{1/2}  \bar H_{1/2 }\big( {s \ov t}\big) 
$$
\eqn\tmmm{
  + \ \sum_{k=2}^\infty d_{k} E_{k- 1/2}(\tau )\ \aa ^{k-1/2} s^k \bar 
H_k\big( {s \ov t} \big)  \ ,\ \ \ \ \ 
g_B={R_{11}\ov R_{10} }\ ,\ \ \aa=R_{10}R_{11}
\ . 
 }
This is manifestly invariant under $SL(2,{\bf Z})$ transformations.
It is interesting to compare \tmmm\ with the amplitude \vex .
The amplitude \tmmm\ should constitute a part of the $\a'\to 0$ limit
of the corresponding non-perturbative 
amplitude in type IIB superstring theory compactified
on the circle $S^1$.
This 
 type IIB amplitude on $S^1$ should be related 
 to the amplitude \vex \ 
in  the decompactification limit $R_{B}\to \infty $ ($R_{10}\to 0$). In this 
limit,
only the second term (proportional to $\zeta(3)E_{3/2}(\tau )$~) survives in
eq. \tmmm , which is the only common term 
of the two amplitudes \vex\ and \tmmm .
It would be very interesting to find the general
type IIB amplitude on $S^1$
from which   eqs.~\vex\ and \tmmm\ can be recovered as special limits. 



For generic values of the coupling, the amplitude
$A_4^{sl(2)}$ given in eq.~\vvx\  has only simple poles.
However, double poles appear when the coupling takes rational values (for example,
if $\tau=ip_0/q_0$, then the masses of the states $(0,1)$,~$n=q_0$ and $(1,0)$,~$n=p_0$
are the same, leading to a double pole in $A_4^{sl(2)}$). In relativistic 
quantum mechanics, double poles are not allowed singularities
of {\it tree}-amplitudes. The meaning of double poles (appearing only for rational coupling)
in the present case  is unclear.
Another delicate point is the interpretation of the residue, which for the amplitude
$A_4^{sl(2)}$ is very complicated to determine.

\smallskip

As in the discussion of ref.~\grgu \ for
$E_{3/2}(\tau )$, one can show that at large $\tau $ 
the functions $E_r(\tau )$ have the following expansion:
\eqn\bbb{
E_r(\tau )=\tau_2^r+\gamma_r \tau_2^{1-r}+
{4\tau_2^{1/2}\pi^r\ov\zeta(2r)\Gamma(r)}
\sum_{n,w=1}^\infty \big({w\ov n}\big)^{r-1/2}\cos(2\pi  wn\tau_1)
K_{r-1/2}(2\pi w n\tau_2 )\ ,
}
$$
\gamma_r={\sqrt{\pi }\ \Gamma(r-1/2)\ \zeta(2r-1)\ov \Gamma(r)\ \zeta(2 r) }\ .
$$
Using the asymptotic expansion for the Bessel function $K_{r-1/2}$,
$$
K_{r-1/2}(2\pi w n\tau_2 )={1\ov \sqrt{4wn\tau_2 } }e^{-2\pi w n\tau_2}\sum_{m=0}^\infty
{1\ov (4\pi wn \tau_2)^m }{\Gamma(r+m)\ov \Gamma(r-m)m! }\ ,
$$
we see that the $E_{k+1/2}(\tau )$ terms in eq. \vex\ are of the form
\eqn\qqq{
g_B^{k+1/2} E_{k+1/2}(\tau )=1+\gamma _{k+1/2}\ g^{2k}_B
+O\big( e^{-2\pi/g_B}\big)\ .
}
The amplitude  \vvx\ thus contains both perturbative and 
non-perturbative contributions. It is important  that all non-perturbative contributions
appear with an exponential factor of the form $e^{-2\pi wn/g_B}$, since
this factor exactly corresponds to the exponential of the 
action of a D-instanton \refs{\grgu,\gv }.
Let us now show that a term of given order in power of momenta will 
only receive a finite number of perturbative contributions.
Using eq.~\qqq , the amplitude $A_4^{ sl(2)}(s,t,u)$ in \vex\ 
can be written as
\eqn\vvv{
\ln A_4^{ sl(2)}(s,t,u)=\ln A_4^0(s,t,u)+ \td A_4(s,t,u)
+O\big( e^{-2\pi/g_B}\big)\ ,
}
with
\eqn\fff{
 \td A_4(s,t,u)=\sqrt{\pi } \sum_{k=1}^\infty
{(k-1)!\zeta (2 k)\ov \Gamma(k+3/2) }g_B^{2k}  
\big(\brs ^{2k+1}+\brt ^{2k+1}+\bru ^{2k+1} \big)\ ,
}
or
\eqn\ccc{
A_4^{ sl(2)}(s,t,u)=A_4^{0}(s,t,u)\big[1+\td A_4 +\ha \td A_4^2+...\big] \ +\ 
O\big( e^{-2\pi/g_B}\big) \ .
}
Thus $A_4^{ sl(2)}$ has the following  structure: 
\eqn\nonp{
A_4^{ sl(2)}(s,t,u)={1\ov \brs\brt\bru }+\sum_{k=0}^\infty
\big[c_0^{(k)}+c_1^{(k)} g^2_B+...+c_{[k/2]+1}^{(k)}  
g^{2[k/2]+2}_B 
+ O\big( e^{-2\pi/g_B}\big)\big]\ \brs ^{k} 
 f_k\big( {s \ov t }\big)\ ,
}
with $c_0^{(1)}=c_1^{(1)}=0\  .$
From eq.~\nonp\ one can read the corrections to higher derivative terms in
the type IIB effective action:
\eqn\efff{
S_{\rm IIB}\bigg|_{\RR^4}
=\int d^{10}x \sqrt{-g} \ 
\sum_{k=0}^\infty 
\big[c_0^{(k)} g_B^{-2}+ c_1^{(k)} +...+c_{[k/2]+1}^{(k)}\  g^{2[k/2]}_B+
O(e^{-2\pi/g_B}) \big] \ (\nabla^2)^{k}\RR^4\  \ .
}
Each term in the sum (containing combinations of $E_{k+1/2}(\tau )$)
 is $SL(2,{\bf Z})$ invariant by itself, as it is explicit in the Einstein frame.
If the amplitude \vex\ were exact, eq.~\efff\ would imply a
non-renormalization theorem. Namely, that
the terms $(\nabla^2)^{2n-2}\RR^4 $ and $(\nabla^2)^{2n-1}\RR^4 $ in the 
$D=10$ type IIB effective action
do not receive perturbative contributions beyond genus $n$.
For $n=1$, this reduces to the observation of \refs{\grgu,\gv } that
$\RR^4 $ should only receive a genus one contribution, but for higher $n$ there
seems to be no reason to expect a non-renormalization theorem to hold.
Terms in type II effective action
that might not receive corrections beyond some 
specific loop order were discussed in
\tte;  examples of such terms are known in
$N=2$ supersymmetric compactifications of type II string theory \ber 
\ (for a discussion of perturbative corrections to
higher derivative terms in the 10D type IIA effective action,
see \rutse ).

Let us now comment on the possible derivation of the amplitude \vex  .
The perturbative part contains higher genus contributions at any order.
It seems hopeless to compute such contributions in a direct way.
For the non-perturbative part $O(e^{-2\pi/g_B})$, a possible approach is
to consider the contribution of D-strings propagating in a genus one
diagram. This was implemented in \gv \ to justify the first term
proportional to $E_{3/2}(\tau )$. One starts with type IIB superstring theory
on $T^2$,
so that a D-instanton of charge $n$ is T-dual to a $(0,n)$ D-string.
Using the euclidean Dirac-Born-Infeld action and going to the Schild gauge
(where an auxiliary field is introduced so that
the action becomes the square of the Nambu action),
the calculation of the amplitude reduces to an ordinary genus one calculation
of string theory with an extra integral over the 
constant part of the auxiliary field.
 In trying to generalize this calculation to the next $E_{k+1/2}(\tau ) $ terms,
 the problem one meets is the integral over the torus moduli space, which
 is now non-trivial.
Nevertheless, the simplicity of the amplitude \vvx\  
 indicates that there may be  another formulation  
 where its determination is more direct. 

Although the amplitude $A_4^{sl(2)}$ seems to be accounting for  contributions
due to exchange of $(p,q)$ strings, it is not clear  that
it can be interpreted as a direct 
sum of string amplitudes with tension $(p,q)$. 
From the point of view of eleven-dimensional supergravity, the 
oscillations  of the different
$(p,q)$ strings correspond to  oscillations of a single
membrane  along the different directions of the 2-torus
(this is explicitly seen from the classical solutions \rusty ).  
When $g_B=R_{11}/R_{10}$ is of order $O(1)$, there is no reason to prefer one particular
set of states (e.g. associated with $(1,0)$ direction) and 
not to include all these  inequivalent states at the same 
time. 
The $SL(2,{\bf Z})$-invariant amplitude constructed here 
naturally incorporates these states and what is known about the 
four-graviton amplitude in type IIB superstring theory
($g_B\to 0$ limit, the D-instanton contributions 
calculated in [3,4], etc.).
If correct, it may provide useful information for 
finding the quantum theory in eleven dimensions.

\bigskip
\bigskip

\noindent{\bf Acknowledgements}
\medskip
I would like to thank  M. Green and A. Tseytlin for useful remarks.
I also  wish to thank SISSA for  hospitality, and  
to  acknowledge the support  of the European
Commission TMR programme grant  ERBFMBI-CT96-0982.
\vfill\eject
\listrefs
\end